\documentstyle[12pt,fleqn]{article}
\def\thebibliography#1{\section*{References}\list
  {[\arabic{enumi}]}{\settowidth\labelwidth{#1}\leftmargin\labelwidth
    \advance\leftmargin\labelsep
    \usecounter{enumi}}
    \def\newblock{\hskip .11em plus .33em minus .07em}
    \sloppy\clubpenalty4000\widowpenalty4000
    \sfcode`\.=1000\relax}

\def\op#1{\mathop{{\it\fam0} #1}\limits}

\newcommand{\id}{{\rm Id\,}}
\newcommand{\Ker}{{\rm Ker\,}}
\newcommand{\beq}{\begin{equation}}
\newcommand{\eeq}{\end{equation}}
\newcommand{\ben}{\begin{eqnarray}}
\newcommand{\een}{\end{eqnarray}}
\newcommand{\be}{\begin{eqnarray*}}
\newcommand{\ee}{\end{eqnarray*}}
\newcommand{\bea}{\begin{eqalph}}
\newcommand{\eea}{\end{eqalph}}
\newcommand{\bR}{{\bf R}}
\newcommand{\rrq}{{\ol q}}
\newcommand{\al}{\alpha}

\newcommand{\la}{\lambda}

\newcommand{\m}{\mu}

\newcommand{\g}{\gamma}
\newcommand{\G}{\Gamma}

\newcommand{\si}{\sigma}
\newcommand{\Si}{\Sigma}

\newcommand{\wt}{\widetilde}
\newcommand{\wh}{\widehat}
\newcommand{\ol}{\overline}
\newcommand{\dr}{\partial}

\newcommand{\ar}{\op\longrightarrow}

\newcommand{\ot}{\otimes}

\newcounter{eqalph}
\newcounter{equationa}
\newcounter{theorem}
\newcounter{proposition}
\newcounter{lemma}
\newcounter{corollary}
\newcounter{definition}

\newenvironment{eqalph}{\stepcounter{equation}
\setcounter{equationa}{\value{equation}}
\setcounter{equation}{0}

\begin{eqnarray}}{\end{eqnarray}\setcounter{equation}{\value{equationa}}}

\def\thedefinition{\arabic{definition}}

\newenvironment{proof}{\noindent 
{\it Proof.}}{\medskip}

\newenvironment{theo}{\refstepcounter{definition} 
\bigskip\noindent{\it Theorem \thedefinition.}}{\medskip}
\newenvironment{prop}{\refstepcounter{definition} 
\bigskip\noindent{\it Proposition \thedefinition.}}{\medskip}
\newenvironment{lem}{\refstepcounter{definition} 
\bigskip\noindent{\it Lemma \thedefinition.}}{\medskip}
\newenvironment{cor}{\refstepcounter{definition} 
\bigskip\noindent{\it Corollary \thedefinition.}}{\medskip}
\newenvironment{defi}{\refstepcounter{definition} 
\bigskip\noindent{\it Definition \thedefinition.}}{\medskip} 

\begin{document}

\hbox{}

{\parindent=0pt 

{ \Large \bf Dynamic connections in analytical mechanics}
\bigskip

{\sc Luigi Mangiarotti$\dagger$\footnote{E-mail
address: mangiaro@camserv.unicam.it} and Gennadi 
Sardanashvily$\ddagger$\footnote{E-mail address: sard@grav.phys.msu.su}}
\medskip

\begin{small}
$\dagger$ Department of Mathematics and Physics, University of Camerino, 62032
Camerino (MC), Italy \\
$\ddagger$ Department of Theoretical Physics, Physics Faculty, Moscow State
University, 117234 Moscow, Russia
\bigskip

{\bf Abstract.} It is shown
that any dynamic equation on a configuration bundle $Q\to\bR$ of
non-relativistic time-dependent mechanics 
is associated with connections on the affine jet bundle $J^1Q\to Q$ and on
the tangent bundle $TQ\to Q$. As a consequence, every non-relativistic
dynamic equation
 can be seen as a geodesic equation with respect to a
(non-linear) connection on the tangent bundle
$TQ\to Q$. Using this fact, the relationship between
relativistic and  non-relativistic equations of motion is studied. The
geometric notions of 
reference frames and  relative accelerations in non-relativistic mechanics are
phrased in the terms of connections. The  covariant form of
non-relativistic dynamic equations is written.
\end{small}
}

\section{Introduction}

We are concerned with non-relativistic time-dependent mechanics whose
configuration space is a bundle $Q\to\bR$ with an $m$-dimensional typical fibre
$M$ over a 1-dimensional base 
$\bR$, treated as a time axis. This configuration space is provided with
bundle coordinates
$(t,q^i)$. The corresponding velocity phase space is the first order jet
manifold $J^1Q$ of sections of the bundle $Q\to\bR$
[2-6,9]. It is coordinated by
$(t,q^i,q^i_t)$.

As is well known, a second order dynamic equation on a
bundle $Q\to\bR$ is defined as a first order dynamic equation on the jet
manifold $J^1Q$, given by a holonomic connection $\xi$ on $J^1Q\to \bR$. The
fact that
$\xi$ is a holonomic curvature-free connection places a limit on the geometric
analysis of dynamic equations.

We aim to show that every dynamic equation on a
configuration space
$Q$  defines a connection $\g_\xi$ on the affine jet bundle $J^1Q\to Q$, and
{\it vice versa}. Then, every dynamic equation on $Q$ can be associated with a
(non-linear) connection
$K$ on the tangent bundle
$TQ\to Q$, and {\it vice versa}. Moreover, 
it gives rise to an equivalent geodesic equation on $TQ$ with respect to
an above-mentioned connection $K$ due to the canonical imbedding $J^1Q\to
TQ$.

In particular, let $Q=X^4$ be a world manifold of a relativistic theory. 
  An equation of motion of a
relativistic system is a geodesic equation 
on  the tangent bundle
$TX$ of relativistic velocities. Thus, both relativistic and non-relativistic
equations of motion can be seen on the tangent bundle $TX$, but their
solutions live in the different subbundles of $TX$. We make use of this fact
in order to study the relationship between relativistic and non-relativistic
equations of motion.

The geometric analysis of dynamic equations also involves  the connections $\G$
on the bundle
$Q\to\bR$ which describe reference frames in non-relativistic mechanics
\cite{eche95,book,book98,sard98}. In particular, one can think of the vertical
vectors
$(q^i_t-\G^i)\dr_i$ on $Q\to\bR$ as being the relative velocities with respect
to the reference frame $\G$. The notion of a relative acceleration is more
intricate. Given a dynamic equation
$\xi$, we define a frame connection $\g_\G$ on $J^1Q\to Q$ and then the lift
$\xi_\G$ of a reference frame
$\G$ to a holonomic connection on
$J^1Q\to\bR$ such that the vertical vector field $a_\G=\xi-\xi_\G$
describes an observable relative acceleration (or a relative force) with
respect to the reference frame $\G$. Then, any dynamic equation can be
written in the form, covariant under coordinate transformations,
\be
\wt D_{\g_\G}q^i_t= a_\G^i
\ee
where $\wt D_{\g_\G}$ is the vertical covariant differential with respect to
the fame connection $\g_\G$.

Throughout the article, the notation 
$\dr/\dr q^\la=\dr_\la$, $\dr/\dr \dot q^\la=\dot \dr_\la$ is used. 

\section{Fibre bundles over $\bR$}

In this inerlude, we point out several important peculiarities of bundles
 over $\bR$. 
The base $\bR$ of $Q\to\bR$ is parameterized by a Cartesian coordinate $t$ 
 with the transition functions
$t'=t+$const. Hence, $\bR$ is
provided with the standard vector field $\dr_t$ and the standard
1-form $dt$. 
The symbol $dt$ also stands for a pull-back of $dt$ onto $Q$.

Any fibre
bundle over $\bR$ is obviously trivial. Every trivialization  
\beq
\psi:Q\cong  \bR\times M \label{gm219}
\eeq
yields the corresponding
trivialization of the jet bundle
\beq
J^1Q\cong \bR\times TM, \qquad \dot q^i=q^i_t. \label{jp2}
\eeq

There is the canonical imbedding 
\ben
&& \la: J^1Q\hookrightarrow TQ, \label{z260}\\
&& \la: (t,q^i,q^i_t) \mapsto (t,q^i,\dot t=1, \dot q^i=q^i_t), \qquad
\la=d_t=\dr_t +q^i_t\dr_i, \nonumber
\een 
where $d_t$ denotes the total derivative.
From now on, we will identify the jet manifold $J^1Q$ with its image in
$TQ$. 

The affine jet bundle
$J^1Q\to Q$ is modelled over the vertical
tangent bundle
$VQ$ of $Q\to\bR$.  As a consequence, we have the canonical
splitting 
\be
\al:V_QJ^1Q\cong J^1Q\op\times_Q VQ, \qquad \al(\dr_i^t)=\dr_i, 
\ee
of the vertical tangent bundle $V_QJ^1Q$ of the affine
jet bundle
$J^1Q\to Q$.
Then the exact sequence  of vector bundles over the
composite bundle $J^1Q\to Q\to\bR$ (see (\ref{63a}) below) reads
\be
&& \put(173,-11){$\rule{0.1mm}{4mm}$}
\put(50,0){\vector(0,-1){10}$\rule{18mm}{0.1mm}\,{}_{\al^{-1}} \,
\rule{18mm}{0.1mm}$} \\
&&0\ar V_QJ^1Q\op\hookrightarrow^i VJ^1Q \ar^{\pi_V} J^1Q\op\times_Q VQ\ar 0.
\ee
Hence, we obtain the linear endomorphism 
\be
 \wh v=i\circ\al^{-1}\circ\pi_V: VJ^1Q\op\to_{J^1Q} VJ^1Q, \qquad 
\wh v\circ \wh v=0,
\ee
of the
vertical tangent bundle $VJ^1Q$ of the jet bundle $J^1Q\to\bR$.
This endomorphism can be extended to the tangent bundle $TJ^1Q$ as follows:
\beq
\wh v(\dr_t) = -q^i_t\dr_i^t,  \qquad \wh v(\dr_i)=\dr^t_i, \qquad
\wh v(\dr^t_i)=0. \label{a1.7}
\eeq

Due to the
monomorphism $\la$ (\ref{z260}), any connection
\beq
\G=dt\ot (\dr_t +\G^i\dr_i) \label{z270}
\eeq
on a fibre bundle $Q\to\bR$ is identified with a nowhere
vanishing  horizontal vector field 
\beq
\G = \dr_t + \G^i \dr_i \label{a1.10}
\eeq
on $Q$. This is the horizontal lift of the
standard vector field $\dr_t$ on $\bR$ by means of the connection 
(\ref{z270}). Conversely, any vector field $\G$ on $Q$ such that
$dt\rfloor\G =1$ defines a connection on $Q\to\bR$. 
Accordingly, the covariant differential associated with a
connection $\G$ on $Q\to\bR$ takes its values into the vertical
tangent bundle of $Q\to\bR$:
\be
 D_\G: J^1Q\op\to_Q VQ, \qquad 
\dot q^i\circ D_\G =q^i_t-\G^i. 
\ee

\begin{prop}\label{gena113}  \cite{book,sard98}.
 Each connection $\G$ on
a bundle $Q\to\bR$ defines an atlas of local constant trivializations of
$Q\to\bR$ such that $\G=\dr_t$ 
with respect to the proper coordinates, and {\it vice versa}. 
In particular, there is one-to-one correspondence between the complete
connections
$\G$ on
$Q\to\bR$ and the trivializations of this bundle.
\end{prop}

 Let $J^1J^1Q$ be the repeated jet manifold of a bundle $Q\to\bR$. It is
coordinated by  $(t,q^i,q^i_t, q^i_{(t)},q^i_{tt})$. 
There are two affine fibrations 
\be
&& \pi_{11}:J^1J^1Q\to J^1Q, \qquad q_t^i\circ\pi_{11} = q_t^i,\\
&& J^1\pi^1_0:J^1J^1Q\to J^1Q,\qquad q_t^i\circ J^1\pi_0^1 =
q_{(t)}^i.
\ee
They are isomorphic by the automorphism $k$ of $J^1J^1Q$ such that
\beq
 q^i_t\circ k=q^i_{(t)}, \qquad q^i_{(t)}\circ k=q^i_t, \qquad
q^i_{tt}\circ k= q^i_{tt}. \label{gm215}
\eeq
The underlying vector bundle of the affine bundle 
$J^1J^1Q\to J^1Q$ is $VJ^1Q\cong J^1VQ$.  

By $J^1_QJ^1Q$ is meant the first order jet manifold of the affine
jet bundle
$J^1Q\to Q$. The adapted coordinates on $J^1_QJ^1Q$ are
$(q^\la,q^i_t,q^i_{\la t})$,  where we use the compact notation
$(q^{\la=0}=t, q^i)$. 

The second order jet manifold $J^2Q$ of a bundle $Q\to\bR$ is coordinated by 
$(t,q^i,q^i_t,q^i_{tt})$. 
The affine bundle $J^2Q\to J^1Q$ is modelled over
the vector bundle
\beq
J^1Q\op\times_QVQ\to J^1Q. \label{gm217}
\eeq
There are the imbeddings
\ben
&& J^2Q \ar^{\la_2} TJ^1Q \ar^{T\la}  V_QTQ\cong T^2Q\subset TTQ, \nonumber
\\ && \la_2: (t,q^i,q^i_t,q^i_{tt})\mapsto (t,q^i,q^i_t,\dot
t=1,\dot q^i=q^i_t,\dot q^i_t=q^i_{tt}). \label{gm211} \\
&& T\la\circ \la_2:(t,q^i,q^i_t,q^i_{tt})\mapsto
 (t,q^i, \dot t =\op t^\circ=1,
\dot q^i =\op q^\circ{}^i=q^i_t, \ddot t=0, \ddot q^i=q^i_{tt}),
\label{cqg80}
\een 
where $(t,q^i,\dot t,\dot q^i,\op t^\circ,\op q^\circ{}^i, \ddot t, \ddot
q^i)$ are the holonomic coordinates on $TTQ$, $V_QTQ$ is the vertical
tangent bundle of $TQ\to Q$, and
$T^2Q$ is a subbundle of
$TTQ$, given by the coordinate relation $\dot t =\op t^\circ$.

Due to the morphism (\ref{gm211}), a connection $\xi$ on the jet bundle
$J^1Q\to
\bR$ is represented by a horizontal vector field on $J^1Q$ such that
$\xi\rfloor dt=1$. 
A connection $\xi$
on $J^1Q\to \bR$ is said to be 
holonomic if it takes its values into $J^2Q$. 

Any connection
$\G$ (\ref{a1.10}) on a bundle $Q\to\bR$ gives rise to
the section $J^1\G$ of the affine bundle $J^1\pi_0^1$ and, 
by virtue of the
isomorphism
$k$ (\ref{gm215}),  to the 
connection
\beq
 J^1\G=\dr_t +\G^i\dr_i + d_t\G^i\dr^t_i \label{gm217'}
\eeq
on the jet bundle $J^1Q\to \bR$.

Here, we also summarize the relevant material on composite bundles (see
\cite{book,sardhp} for details).
Let us consider the composite bundle
\beq
 Y\to \Si\to X, \label{1.34}
\eeq
where $Y\to\Si$  and $\Si\to X$
are bundles. It is
equipped with bundle coordinates $(x^\la,\si^m,y^i)$ where
$(x^\m,\si^m)$ are bundle coordinates on the bundle $\Si\to X$ such that the
transition functions $\si^m\to\si'^m(x^\la,\si^k)$ are independent of the
coordinates $y^i$.

Let us consider the jet manifolds $J^1\Si$,
$J^1_\Si Y$ and $J^1Y$ of the bundles
$\Si\to X$, $Y\to \Si$ and $Y\to X$, respectively. They are coordinated by
\be
( x^\la ,\si^m, \si^m_\la),\quad 
( x^\la ,\si^m, y^i, \wt y^i_\la, y^i_m),\quad 
( x^\la ,\si^m, y^i, \si^m_\la ,y^i_\la) .
\ee
We have the following canonical map \cite{sau}:
\beq
\rho: J^1\Si\op\times_\Si J^1_\Si Y\ar_Y J^1Y, 
\qquad y^i_\la\circ\rho=y^i_m{\si}^m_{\la} +\wt y^i_{\la}.\label{1.38}
\eeq

Given a composite bundle $Y$ (\ref{1.34}), we have the exact
sequence 
\beq
 0\to V_\Si Y\hookrightarrow VY\to Y\op\times_\Si V\Si\to 0, \label{63a}
\eeq
where $V_\Si Y$ is the vertical tangent bundle of 
$Y\to\Si$.
Every connection 
\beq
A_\Si=dx^\la\ot(\dr_\la+\wt A^i_\la\dr_i)
+ d\si^m\ot(\dr_m+A^i_m\dr_i) \label{Q1}
\eeq
 on $Y\to\Si$ determines
the splitting
\be
&& VY=V_\Si Y\op\oplus_Y A_\Si(Y\op\times_\Si V\Si),\\
&& \dot y^i\dr_i + \dot\si^m\dr_m=
(\dot y^i -A^i_m\dot\si^m)\dr_i + \dot\si^m(\dr_m+A^i_m\dr_i), 
\ee
of the exact sequence (\ref{63a}).
Using this splitting, one can construct the first order
differential operator, called the vertical covariant differential,  
\beq
\wt D: J^1Y\to T^*X\op\otimes_Y V_\Si Y,
\quad \wt D= dx^\la\otimes(y^i_\la-\wt
A^i_\la -A^i_m\si^m_\la)\dr_i, \label{7.10}
\eeq
on the composite bundle
$Y\to X$.

Given a connection $A_\Si$ (\ref{Q1}) on the bundle $Y\to\Si$ and a
connection 
\be
B=dx^\la\ot(\dr_\la + B^m_\la\dr_m +B^i_\la\dr_\la)
\ee
on the composite bundle $Y\to X$, there exists another connection
$A'_\Si$ on the bundle $Y\to\Si$ with the components 
\beq
 A'^i_m=  A^i_m, \qquad  A'^i_\la=B^i_\la- A^i_m B^m_\la. \label{jp10}
\eeq

\section{Equations on a manifold}

Let $N$ be a manifold, coordinated by $(q^\la)$. We recall some notions. 

\begin{defi}
A second order equation
on a manifold
$N$ is said to be an image $\Xi(TN)$ of a holonomic vector field
\be
\Xi=\dot q^\la \dr_\la + u^\la\dot\dr_\la 
\ee
on the tangent bundle $TN$. It is a closed imbedded subbundle of 
$TTN\to N$, given by the coordinate conditions
\beq
\op q^\circ{}^\la=\dot q^\la, \qquad \ddot q^\la=u^\la(q^\m,\dot q^\m).
\label{jp11}
\eeq
\end{defi}

By a  solution of a second order equation on $N$ is meant a curve $c:()\to N$
whose  second order tangent prolongation $\ddot c$ lives in the subbundle
(\ref{jp11}).

Given a connection 
\beq
K= dq^\la\ot (\dr_\la +K^\m_\la\dot\dr_\m) 
\label{z290}
\eeq
on the tangent bundle $TN\to N$, let
\beq
\wh K: TN\op\times_N TN\to TTN \label{jp21}
\eeq
be the corresponding linear bundle morphism
over $TN$ which splits the exact sequence
\be
0\ar V_NTN\hookrightarrow TTN\ar TN\op\times_N TN\ar 0.
\ee

\begin{defi}
A geodesic equation on $TN$ with respect to the
connection $K$ is defined as the image
\beq
 \op q^\circ{}^\m=\dot q^\m, \qquad  \ddot q^\m = K^\m_\la\dot
q^\la \label{jp20}
\eeq
of the morphism (\ref{jp21}) restricted to the diagonal $TN\subset TN\times
TN$. 
\end{defi}

By a solution of a geodesic equation on $TN$ is meant a geodesic curve $c:()\to
N$, whose tangent prolongation $\dot c$ is an integral
section (a geodesic vector field) over $c\subset N$ for the connection $K$.
The geodesic equation
(\ref{jp20}) can be written in the form
\be
\dot q^\la\dr_\la\dot q^\m=K^\m_\la\dot
q^\la, 
\ee
where by $\dot q^\m(q^\al)$ is meant a geodesic vector field (which exists at
least on a geodesic curve), while $\dot q^\la\dr_\la$ is a formal operator of
differentiation (along a curve).

It is readily observed that the morphism $\wh K\mid_{TN}$
is a holonomic vector field on $TN$. It follows that any geodesic equation
(\ref{jp21}) on $TN$ is a second order equation on $N$. The converse is
not true in general. Nevertheless, we have the following theorem.

\begin{theo} \label{jp30}  \cite{mora}. Every second order equation
(\ref{jp11}) on a manifold $N$ defines a connection $K_\Xi$ on the tangent
bundle
$TN\to N$ whose components are
\beq
K^\m_\la = \frac12\dot\dr_\la \Xi^\m. \label{jp32}
\eeq
 \end{theo}

However, the second order equation (\ref{jp11}) fails
to be a geodesic equation with respect to the connection (\ref{jp32}) in
general. In particular, the geodesic equation (\ref{jp20}) with respect to a
connection
$K$ determines the connection (\ref{jp32}) on $TN\to N$ which does not
necessarily coincide with $K$. A second order equation $\Xi$ on $N$ is a
geodesic equation for the connection (\ref{jp32}) if and
only if
$u$ is a spray, i.e., $[v,\Xi]=\Xi$,
where $v=\dot q^\la\dot\dr_\la$ is the Liouville vector field on $TN$. In
Section 5, we will improve Theorem \ref{jp30}.

\section{Dynamic equations}

Let $Q\to\bR$ be a bundle coordinated by $(t,q^i)$.

\begin{defi} \label{gena73}
A second order differential equation on $Q\to\bR$, called a dynamic
equation,  is defined as the image $\xi(J^1Q)\subset J^2Q$ 
of a holonomic connection
\beq
\xi =\dr_t +q^i_t\dr_i + \xi^i(t,q^j,q^j_t)\dr_i^t \label{jp25}
\eeq 
on $J^1Q\to\bR$. This is 
a closed
subbundle of 
$J^2Q\to \bR$,
 given by the coordinate relations
\beq
q^i_{tt}=\xi^i(t,q^j,q^j_t). \label{z273}
\eeq
\end{defi}

A solution of the dynamic
equation (\ref{z273}), called a motion, is a curve $c:()\to Q$  whose
second order jet prolongation $J^2c:()\to J^1Q$ lives in (\ref{z273}).

One can easily find the transformation law 
\beq
q'^i_{tt} = \xi'^i, \qquad \xi'^i=(\xi^j\dr_j + q^j_tq^k_t\dr_j\dr_k
+2q^j_t\dr_j\dr_t +\dr_t^2)q'^i(t,q^j) \label{z317}
\eeq
of a dynamic equation under
coordinate transformations $q^i\to q'^i(t,q^j)$.

A dynamic equation $\xi$ on a bundle $Q\to \bR$ is said to be conservative if
there exists a trivialization (\ref{gm219}) of $Q$ and the
corresponding trivialization (\ref{jp2}) of $J^1Q$ such that the vector field 
$\xi$ (\ref{jp25}) on $J^1Q$ is projectable onto $M$. Then this projection
\be
\Xi_\xi=\dot q^i\dr_i +\xi^i(q^j,\dot q^j)\dot \dr_i 
\ee
is a second order equation on the typical fibre $M$ of $Q$. Conversely, every
second order equation $\Xi$ on a manifold $M$ can be seen as a
conservative dynamic equation
\beq
\xi_\Xi=\dr_t + \dot q^i\dr_i + u^i\dot \dr_i \label{jp27}
\eeq
on the bundle $\bR\times M\to\bR$ in accordance with the isomorphism
(\ref{jp2}).

\begin{prop}\label{jp28} 
Any dynamic equation on a bundle $Q\to\bR$ is equivalent to a second order
equation on a manifold $Q$.
\end{prop}

\begin{proof}
Given a dynamic equation $\xi$ on a bundle $Q\to\bR$, let us
consider the diagram
\beq
\begin{array}{rcccl}
& J^2Q & \ar & T^2Q & \\
_\xi &  \put(0,-10){\vector(0,1){20}} & &  \put(0,-10){\vector(0,1){20}}
& _\Xi\\
& J^1Q &\ar^\la & TQ &
\end{array} \label{jp14}
\eeq
where $\Xi$ is a holonomic vector field on $TQ$, and we use the
morphism (\ref{cqg80}). A glance at the expression (\ref{cqg80}) shows that
the  diagram (\ref{jp14}) can be commutative only if the component $\Xi^0$ of a
vector field $\Xi$ vanishes. Since the transition functions $t\to
t'$ are independent of $q^i$, such a vector field may exist on $TQ$.
Now the diagram (\ref{jp14}) becomes commutative if the dynamic equation
$\xi$ and a vector field $\Xi$ fulfill the relation 
\beq
\xi^i=\Xi^i(t,q^j,\dot t=1, \dot q^j=q^j_t).\label{jp29}
\eeq
It is easily seen that this relation holds globally because the substitution
of $\dot q^i=q^i_t$ into the transformation law of a vector field $\Xi$
restates the transformation law (\ref{z317}) of the holonomic connection
$\xi$. In accordance with the relation (\ref{jp29}), a desired vector field
$\Xi$ is an extension  of the section $T\la\circ\la_2\circ \xi$ 
of the bundle $TTQ\to TQ$ over the closed submanifold $J^1Q\subset TQ$ to a
global section. Such an extension always exists, but is not unique. Then, the
dynamic equation (\ref{z273}) can be written in the form
\beq
q^i_{tt}= \Xi^i\mid_{\dot t=1, \dot q^j=q^j_t}. \label{jp36}
\eeq
It is equivalent to the
second order equation on
$Q$
\beq
\ddot t=0, \qquad \dot t=1, \qquad \ddot q^i= \Xi^i. \label{jp35}
\eeq
 Being a solution of (\ref{jp35}), a curve $c$ in $Q$ also fulfills
(\ref{jp36}), and {\it vice versa}.  
\end{proof}

It should be emphasized that, written in the bundle coordinates
$(t,q^i)$, the second order equation  (\ref{jp35}) is well defined with respect
to any coordinates on $Q$.

\section{Dynamic connections}

To say more than Proposition \ref{jp28}, we turn  to  the relationship
between the dynamic equations on $Q$ and the connections
on the affine jet bundle $J^1Q\to Q$. 
Let
\beq
\g:J^1Q\to J^1_QJ^1Q, \qquad  \g=dq^\la\ot (\dr_\la + \g^i_\la \dr_i^t),
\label{a1.38}
\eeq
be such a connection. Its coordinate
transformation law is
\beq
 \g'^i_\la = (\dr_jq'^i\g^j_\m
+\dr_\m q'^i_t)\frac{\dr q^\m}{\dr q'^\la}. \label{m175}
\eeq

\begin{prop}\label{gena51}
Any connection $\g$ (\ref{a1.38}) on the affine jet bundle $J^1Q\to Q$ defines
the holonomic connection
\beq
 \xi_\g = \dr_t + q^i_t\dr_i +(\g^i_0
+q^j_t\g^i_j)\dr_i^t,\label{z281}
\eeq
on the jet bundle
$J^1Q\to \bR$.  
\end{prop}

\begin{proof}
Let us consider the composite bundle  
$J^1Q\to Q\to\bR$
and the canonical morphism $\rho$ (\ref{1.38}) 
 which reads
\beq
\rho:  J^1_QJ^1Q \ni (q^\la,q^i_t,q^i_{\la t}) \mapsto
(q^\la,q^i_t,q^i_{(t)}=q^i_t,q^i_{tt}=q^i_{0t} +q^j_tq^i_{jt})\in J^2Q.
\label{z298}
\eeq
A connection $\g$ (\ref{a1.38}) and the morphism $\rho$ (\ref{z298})
combine into the desired holonomic connection  $\xi_\g$ (\ref{z281})
on the jet bundle $J^1Q\to\bR$.
\end{proof}

It follows that each connection $\g$ (\ref{a1.38}) on the affine jet bundle 
$J^1Q\to Q$ yields the dynamic equation
\beq
q^i_{tt}=(\g^i_0 +q^j_t\g^i_j) \label{z287}
\eeq
on the bundle $Q\to\bR$. 
This is exactly the restriction to $J^2Q$ of the kernel
$\Ker \wt D_\g$ of the vertical covariant differential $\wt D_\g$
(\ref{7.10}) defined by the connection $\g$:
\be
 \wt D_\g: J^1J^1Q\to V_QJ^1Q,\qquad 
\dot q^i_t\circ\wt D_\g= q^i_{tt} -\g^i_0 - q^j_t\g^i_j.
\ee
Therefore, connections on
$J^1Q\to Q$ are also called dynamic connections (one should distinguish
this terminology from that of
\cite{massa}). Of course, different dynamic connections may lead to the same
dynamic equation (\ref{z287}).

\begin{prop}\label{gena52}
Any holonomic connection $\xi$ (\ref{jp25}) on the jet bundle
$J^1Q\to \bR$ yields the dynamic connection 
\beq
\g_\xi =dt\ot[\dr_t+(\xi^i-\frac12 q^j_t\dr_j^t\xi^i)\dr_i^t] +
dq^j\ot[\dr_j +\frac12\dr_j^t\xi^i \dr_i^t]
\label{z286}
\eeq
 on the affine
jet bundle $J^1Q\to Q$.
\end{prop}

\begin{proof} Given an arbitrary
vector field  $u = a^i\dr_i + b^i\dr_i^t$
on the jet bundle $J^1Q\to\bR$, let us put
\be
I_\xi (u) = [\xi,\wh v(u)] - \wh v([\xi,u])= -a^i \dr_i + (b^i -
a^j\dr^t_j\xi^i)\dr_i^t,
\ee
where $\wh v$ is the endomorphism (\ref{a1.7}). We come to 
the endomorphism 
\be
&& I_\xi: VJ^1Q\op\to_{J^1Q} VJ^1Q, \\
&& I_\xi: \dot q^i\dr_i +\dot q^i_t\dr_i^t \mapsto -\dot q^i\dr_i
+(\dot q^i_t-\dot q^j \dr^t_j\xi^i) \dr_i^t, 
\ee
which obeys the condition  $I_\xi\circ I_\xi=I_\xi$.
Then there is the projection
\be
&&J_\xi=\frac12(I_\xi +\id VJ^1Q):VJ^1Q\op\to_{J^1Q} V_QJ^1Q,\\
&&J_\xi:\dot q^i\dr_i +\dot q^i_t\dr^t_i \mapsto (\dot q^i_t-\frac12\dot q^j
\dr^t_j\xi^i) \dr_i^t.
\ee
Recall that a holonomic connections $\xi$ on $J^1Q\to\bR$ defines
the projection 
\be
\wh\xi: TJ^1Q\ni \dot t\dr_t + \dot q^i\dr_i +\dot q^i_t\dr^t_i \mapsto 
(\dot q^i -\dot tq^i_t)\dr_i + (\dot q^i_t -\dot \xi^i)\dr_i^t\in VJ^1Q.
\ee
Then the composition 
\be
&& I_\xi\circ \wh\xi: TJ^1Q \to VJ^1Q\to V_QJ^1Q,\\
&& \dot t\dr_t + \dot q^i\dr_i +\dot q^i_t\dr^t_i \mapsto 
[\dot q^i_t- \dot t(\xi^i-\frac12 q^j_t\dr_j^t\xi^i) -\frac12\dot q^j
\dr_j^t\xi^i ]\dr_i^t,
\ee
corresponds to the connection $\g_\xi$ (\ref{z286})
on the affine jet bundle $J^1Q\to Q$.
\end{proof}

The dynamic connection $\g_\xi$ (\ref{z286}) possesses the property
\be
\g^k_i = \dr_i^t\g^k_0 +  q^j_t\dr_i^t\g^k_j
\ee
which implies $\dr_j^t\g^k_i = \dr_i^t\g^k_j$. 
Such a dynamic connection is called symmetric.

Let $\g$ be a dynamic connection (\ref{a1.38}) and $\xi_\g$ the 
corresponding dynamic equation (\ref{z281}). Then 
the dynamic connection associated with $\xi_\g$ takes the form
\be
\g_{\xi_\g}{}^k_i = \frac{1}{2}
(\g^k_i + \dr_i^t\g^k_0 + q^j_t\dr_i^t\g^k_j),
\qquad \g_{\xi_\g}{}^k_0 = \xi^k - q^i_t\g_{\xi_\g}{}^k_i. 
\ee
It is readily observed that $\g = \g_{\xi_\g}$ if and only if $\g$ 
is symmetric.

Since the jet bundle $J^1Q\to Q$ is affine, it admits an affine
connection 
\be
 \g=dq^\la\ot [\dr_\la + (\g^i_{\la 0}(q^\al)+ \g^i_{\la
j}(q^\al)q^j_t)\dr_i^t].
\ee
This connection is symmetric if and only if $\g^i_{\la\m}=\g^i_{\m\la}$. 
An affine dynamic connection generates a quadratic
dynamic equation, and {\it vice versa}. 

We use a dynamic connection in order to modify Theorem
\ref{jp30}. Let $\Xi$ be a second order equation on a manifold $N$ and 
$\xi_\Xi$ (\ref{jp27}) the corresponding conservative dynamic equation on
the bundle $\bR\times N\to\bR$. The latter yields the dynamic connection $\g$
(\ref{z286}) on the bundle 
\be
\bR\times TN\to \bR\times N.
\ee
Its components $\g^\m_\la$ are exactly those of the connection (\ref{jp32}) on
$TN\to N$ from Theorem \ref{jp30}, while $\g^\m_0$ make up 
a vertical vector field
\beq
e=\g^\m_0\dot \dr_\m = (\Xi^\m -\frac12 \dot q^\la\dot\dr_\la \Xi^\m)\dot
\dr_\m
\label{jp38}
\eeq
on $TN\to N$. Thus, we have proved the
following.

\begin{prop}
Every second order equation $\Xi$ (\ref{jp11}) on a manifold $N$ admits the
decomposition
\be
\Xi^\m= K^\m_\la\dot q^\la + e^\m
\ee
where $K$ is the connection (\ref{jp32}) on $TN\to N$, and $e$ is the
vertical vector field (\ref{jp38}).
\end{prop}

With a dynamic connection $\g_\xi$ (\ref{z286}), one can also restate the
linear connection on $TJ^1Q\to Q$, associated with a dynamic equation on
$Q$ \cite{massa} (see \cite{book98} for details).

\section{Non-relativistic geodesic equations}

To improve Proposition \ref{jp28}, we aim to show that every dynamic equation
on a bundle $Q\to\bR$ is equivalent to a geodesic equation on the tangent
bundle $TQ\to Q$.

Let
us consider the diagram
\beq
\begin{array}{rcccl}
& J^1_QJ^1Q & \ar^{J^1\la} & J^1_QTQ & \\
_\g &  \put(0,-10){\vector(0,1){20}} & &  \put(0,-10){\vector(0,1){20}}
& _K\\
& J^1Q &\ar^\la & TQ &
\end{array} \label{z291}
\eeq
where $J^1_QTQ$ is the first order jet manifold of the tangent bundle $TQ\to
Q$, coordinated by $(t,q^i,\dot t,\dot q^i, (\dot t)_\m,
(\dot q^i)_\m)$, while $K$ 
is a connection (\ref{z290}) on $TQ\to Q$.

The jet prolongation over $Q$ of the morphism $\la$
(\ref{z260}) reads 
\be
J^1\la: (t,q^i,q^i_t, q^i_{\m t}) \mapsto 
(t,q^i,\dot t=1,\dot q^i=q^i_t, (\dot t)_\m=0,
(\dot q^i)_\m=q^i_{\m t}).
\ee
We have
\be
&& J^1\la\circ \g: (t,q^i,q^i_t) \mapsto 
(t,q^i,\dot t=1,\dot q^i=q^i_t, (\dot t)_\m=0,
(\dot q^i)_\m=\g^i_\m ),\\
&& K\circ \la: (t,q^i,q^i_t) \mapsto 
(t,q^i,\dot t=1,\dot q^i=q^i_t, (\dot t)_\m=K_\m^0,
(\dot q^i)_\m=K^i_\m).
\ee
It follows that the diagram (\ref{z291}) can be commutative only
if the components $K^0_\m$ of the connection $K$
vanish.  Since the coordinate transition functions $t\to t'$ are independent
of $q^i$, a connection 
\beq
\wt K=dq^\la\ot (\dr_\la +K^i_\la\dot\dr_i)
\label{z292}
\eeq
with $K^0_\m=0$ may exist on $TQ\to Q$. It
obeys the transformation law
\beq
{K'}_\la^i=(\dr_j q'^i K^j_\m + \dr_\m\dot q'^i)
\frac{\dr q^\m}{\dr q'^\la}.  \label{z293}
\eeq
Now the diagram (\ref{z291}) becomes commutative if the connections
$\g$ and $\wt K$ fulfill the relation
\beq
\g^i_\m= K^i_\m\circ \la=K^i_\m(t,q^j,\dot t=1, \dot q^j=q^j_t).
\label{z294}
\eeq
It is easily seen that this relation holds globally because the
substitution of $\dot q^i=q^i_t$ into (\ref{z293}) restates the
transformation law (\ref{m175}) of a connection on the affine jet
bundle $J^1Q\to Q$. 
In accordance with the relation (\ref{z294}), a desired
connection $\wt K$ is an extension  of the section
$J^1\la\circ \g$ of the affine bundle $J^1_QTQ\to TQ$ over the closed
submanifold $J^1Q\subset TQ$ to a global section. Such an extension
always exists, but is not unique. 
Thus, it is stated the following.

\begin{prop} \label{motion1} 
In accordance with the relation (\ref{z294}), every dynamic equation
(\ref{z273}) on the configuration space $Q$ can be written in the form
\beq
q^i_{tt} = K^i_0\circ\la +q^j_t K^i_j\circ\la, \label{gm340}
\eeq
where $\wt K$ is a connection (\ref{z292}). Conversely,
 each connection $\wt K$ (\ref{z292}) on the tangent bundle $TQ\to Q$ defines
a dynamic connection 
$\g$ on the affine jet bundle $J^1Q\to Q$ and the dynamic equation
(\ref{gm340}) on the configuration space $Q$.
\end{prop}

Then we come to the following theorem.

\begin{theo} \label{jp50}
Every dynamic equation
(\ref{z273}) on the configuration space $Q$ is equivalent to the geodesic
equation 
\beq
\ddot t=0, \qquad \dot t=1,\qquad 
 \ddot q^i = K^i_\la\dot q^\la, \label{cqg11}
\eeq
on the tangent bundle $TQ$ relative to a connection $\wt K$ with the
components $K^0_\m=0$ and $K^i_\m$ (\ref{z294}).
Its solution is a geodesic curve in $Q$ which also obeys the dynamic
equation (\ref{gm340}), and {\it vice versa}.
\end{theo} 

In accordance with this theorem,  the second order equation (\ref{jp35}) in
Proposition \ref{jp28} can be chosen as a geodesic equation.
It should be emphasized that, written in the bundle coordinates
$(t,q^i)$, the geodesic equation (\ref{cqg11}) and the connection $\wt K$
(\ref{z294}) are well defined with respect to any coordinates on $Q$. 

From the physical viewpoint, the most interesting dynamic equations are the
quadratic ones
\beq 
\xi^i = a^i_{jk}(q^\m)q^j_t q^k_t + b^i_j(q^\m)q^j_t + f^i(q^\m).
\label{cqg12}
\eeq
This property is global due to the transformation law (\ref{z317}). Then one
can use the following two facts.

\begin{prop}\label{aff}
There is one-to-one correspondence between the affine connections $\g$ on
 $J^1Q\to Q$ and the linear symmetric connections $K$
(\ref{z292}) on $TQ\to Q$. This correspondence is
given by the relation (\ref{z294}) which takes the form
\be
&& \g^i_\m=\g^i_{\m 0} + \g^i_{\m j}q^j_t =K_\m{}^i{}_0(q)\dot t + 
K_\m{}^i{}_j(q)\dot q^j|_{\dot t=1, \dot q^i=q^i_t}=
K_\m{}^i{}_0(q) + 
K_\m{}^i{}_j(q)q^j_t, \\
&& \g^i_{\m\la}= K_\m{}^i{}_\la. 
\ee
\end{prop}

\begin{cor} \label{c2}
Every quadratic dynamic equation (\ref{cqg12})
gives rise to the geodesic equation
\ben
&& \ddot q^0= 0, \qquad \dot q^0=1,\nonumber\\
&& \ddot q^i= 
a^i_{jk}(q^\m)\dot q^j \dot q^k + b^i_j(q^\m)\dot q^j\dot q^0 +
f^i(q^\m) \dot q^0\dot q^0 \label{cqg17}
\een
on $TQ$ with respect to the symmetric linear connection 
\beq
K_\la{}^0{}_\nu=0, \quad K_0{}^\i{}_0= f^i, \quad K_0{}^\i{}_j=\frac12 b^i_j,
\quad K_k{}^i{}_j= a^i_{kj}. \label{cqg13}
\eeq
\end{cor}

The geodesic equation
(\ref{cqg17}) however is not unique for the dynamic equation (\ref{cqg12}).

\begin{prop} \label{jp40} 
Any quadratic dynamic equation (\ref{cqg12}), being 
equivalent to the geodesic equation with respect to the linear connection
$\wt K$ (\ref{cqg13}), is also equivalent to the  one with respect to an affine
connection $K'$ on $TQ\to Q$ which differs from $\wt K$ (\ref{cqg13}) in a
soldering form $\si$ on $TQ\to Q$ with the components
\be
\si^0_\la= 0, \qquad \si^i_k= h^i_k-\frac12 h^i_k\dot x^0, \qquad \si^i_0=
-\frac12 h^i_k\dot x^k -h^i_0\dot x^0 + h^i_0,
\ee
where $h^i_\la$ are local functions on $Q$.
\end{prop} 

Let us extend our inspection of  dynamic equations and connections 
to connections on the tangent bundle $TM\to
M$ of the typical fibre of the configuration space $Q\to \bR$.
In this case, the relationship fails to be  canonical, but depends on a
trivialization (\ref{gm219}).

Given such a trivialization, let $(t,\rrq^i)$ be the associated coordinates on
$Q$, where
$\rrq^i$ are coordinates on $M$ with transition functions independent of
$t$.  The corresponding trivialization (\ref{jp2}) of $J^1Q\to\bR$
takes place in the coordinates $(t,\rrq^i,\dot\rrq^i)$. With respect to these
coordinates, the transformation law (\ref{m175}) of a dynamic connection $\g$
on the affine jet bundle $J^1Q\to Q$ reads 
\be
{\g'}_0^i =\frac{\dr\rrq'^i}{\dr\rrq^j}\g_0^j \qquad
\g'^i_k = (\frac{\dr\rrq'^i}{\dr \rrq^j}\g^j_n
+\frac{\dr \dot \rrq'^i}{\dr \rrq^n})\frac{\dr\rrq^n}{\dr{\rrq'}^k}.
\ee
It follows that, given a trivialization of $Q\to\bR$, a dynamic connection
$\g$ defines the  time-dependent vertical vector
field 
\be
\g^i_0(t,\rrq^j,\dot\rrq^j)\frac{\dr}{\dr \dot\rrq^i}:\bR\times TM\to VTM
\ee
 and the time-dependent connection 
\beq
d\rrq^k\ot\left(\frac{\dr}{\dr \rrq^k} +\g^i_k(t,\rrq^j, \dot\rrq^j)
\frac{\dr}{\dr \dot\rrq^i}\right): \bR\times TM\to J^1TM \subset TTM
\label{jp45}
\eeq
 on the tangent
bundle $TM\to M$.

Conversely, let us consider a connection 
\be
\ol K=d\rrq^k\ot\left(\frac{\dr}{\dr \rrq^k} +\ol K^i_k(\rrq^j, \dot\rrq^j)
\frac{\dr}{\dr \dot\rrq^i}\right) 
\ee
on the tangent bundle $TM\to M$. Given the above-mentioned
trivialization of $Q\to\bR$, the connection $\ol K$
defines the connection $\wt K$ (\ref{z292}) with the components 
\be
K_0^i=0, \qquad K^i_k=\ol K^i_k,
\ee
on the tangent bundle $TQ\to Q$. The corresponding dynamic
connection $\g$ on the affine jet bundle $J^1Q\to Q$ reads
\beq
\ol \g_0^i=0, \qquad \ol \g^i_k=\ol K^i_k. \label{m177}
\eeq
Using the transformation law (\ref{m175}), one can extend the
expression (\ref{m177}) to arbitrary bundle coordinates 
on the configuration space $Q$. 
Thus, we have proved the following.

\begin{prop}\label{jp46}
Any connection $\ol K$ on the typical fibre $M$ of a bundle $Q\to
\bR$ yields a conservative dynamic equation on $Q$.
\end{prop}

\section{Reference frames}

From the physical viewpoint,
a reference frame in non-relativistic mechanics sets a tangent vector at
each point of a configuration space
$Q$ which characterizes the velocity of an "observer" at this point. Thus, we
come to the following geometric definition of a reference frame.

\begin{defi}
In non-relativistic mechanics, a reference frame is said to be a connection
$\G$ on the bundle $Q\to\bR$.
\end{defi}

In accordance with this definition, the corresponding covariant
differential 
\be
D_\G(q^i_t)= q^i_t-\G^i=\dot q^i_\G 
\ee
determines the relative velocities with respect to the reference
frame $\G$.
In particular, given a motion $c$ in $Q$, the
covariant derivative
$\nabla^\G c$ is the velocity of this motion relative to the
reference frame $\G$. For instance, if $c$ is an integral section of
the connection $\G$, the relative velocity of $c$ with respect to the
reference frame $\G$ is equal to 0. Conversely, every motion $c:\bR\to
Q$, defines a proper reference frame $\G_c$  such that the velocity of $c$
relative to $\G_c$ equals 0. This reference frame $\G_c$ is an extension of the
local section $J^1c: c(\bR)\to J^1Q$  of the affine jet bundle $J^1Q\to Q$ 
to a global section. Such a global section always exists.

By virtue of Proposition \ref{gena113}, any reference frame $\G$ on the
configuration space 
$Q\to\bR$ is associated with an atlas  of local constant
trivializations such that $\G=\dr_t$
with respect to the corresponding coordinates $(t,\ol q^i)$ whose transition
functions are independent of time. Such an atlas is also called a reference
frame. A reference frame is said to be complete if the
associated connection $\G$ is complete.  In accordance with Proposition
\ref{gena113} every complete reference frame provides a trivialization  
of a bundle $Q\to\bR$, and {\it vice versa}. 

Using the notion of a reference frame, we obtain a converse of Theorem
\ref{jp50}.

\begin{theo}\label{jp51} 
Given a reference frame $\G$, any connection $K$
(\ref{z290}) on the tangent bundle $TQ\to Q$ defines a dynamic equation
\be
\xi^i= (K^i_\la -\G^i K^0_\la)\dot q^\la\mid_{\dot q^0=1,\dot q^j=q^j_t}.
\ee
\end{theo}

The proof follows at once from Proposition \ref{motion1} and the following
lemma.

\begin{lem}
Given a connection $\G$ on the bundle $Q\to\bR$ and a connection $K$ on the
tangent bundle $TQ\to Q$, there is the connection $\wt K$ on $TQ\to Q$ with
the components
\be
\wt K^0_\la =0, \qquad \wt K^i_\la = K^i_\la - \G^iK^0_\la.
\ee
\end{lem}

Let us point out the following interesting class of dynamic equations which
we agree to call the free motion equations.   

\begin{defi}
We say that the dynamic equation (\ref{z273}) is a free motion
equation if there exists a reference frame $(t,\ol q^i)$ on the configuration
space $Q$ 
 such that this equation reads 
\beq
\ol q^i_{tt}=0. \label{z280}
\eeq
\end{defi}

With respect to arbitrary bundle coordinates
$(t,q^i)$, a free motion equation takes the form
\beq
 q^i_{tt}=d_t\G^i +\dr_j\G^i(q^j_t-\G^j) -
\frac{\dr q^i}{\dr\rrq^m}\frac{\dr\rrq^m}{\dr q^j\dr q^k}(q^j_t-\G^j)
(q^k_t-\G^k),  \label{m188}
\eeq
where $\G^i=\dr_t q^i(t,\ol q^j)$ is the connection associated with the initial
frame $(t,\ol q^i)$.  One can think of the right hand side of the equation
(\ref{m188}) as being the general coordinate expression of an inertial force
in non-relativistic mechanics. The corresponding
dynamic connection $\g$ on the affine jet bundle $J^1Q\to Q$ reads
\be
 \g^i_k=\dr_k\G^i  -
\frac{\dr q^i}{\dr\rrq^m}\frac{\dr\rrq^m}{\dr q^j\dr q^k}(q^j_t-\G^j),
\qquad
\g^i_0= \dr_t\G^i +\dr_j\G^iq^j_t -\g^i_k\G^k. 
\ee
It is affine. In virtue of Proposition \ref{aff}, this dynamic connection
defines a linear connection $K$ on the tangent bundle $TQ\to Q$ whose
curvature is necessarily equal to 0.  Thus, we come to the following
criterion of a dynamic equation to be a free motion equation.

\begin{prop}
If $\xi$ is a free motion
equation on a configuration space $Q$, it is quadratic and the corresponding
linear symmetric connection (\ref{cqg13}) on the tangent bundle $TQ\to Q$ is
a curvature-free connection.
\end{prop}

This criterion fails to be a sufficient condition because it may happen that
the components of a curvature-free  symmetric linear connection on
$TQ\to Q$ vanish with respect to the coordinates on $Q$
 which are not compatible with the fibration $Q\to\bR$. Nevertheless, we can
formulate the necessary and sufficient condition of existence of a free
motion equation on a configuration space $Q$.

\begin{prop} \label{gena110}
A free motion equation on a bundle $Q\to\bR$ exists if and only if
the typical fibre $M$ of $Q$ admits a curvature-free symmetric linear
connection.
\end{prop}

\begin{proof}
Let a free motion equation take the form (\ref{z280}) with respect
to an atlas of bundle coordinates on $Q\to\bR$. By virtue of
Proposition
\ref{gena52}, there exists an affine dynamic connection $\g$ on the affine jet
bundle $J^1Q\to Q$ whose components relative to this atlas 
are equal to 0. Given a trivialization chart of this atlas, this connection
defines the curvature-free symmetric linear connection on $M$ (\ref{jp45}). 
The converse statement follows at once from Proposition  \ref{jp46}.
\end{proof}

\section{Relative acceleration}

To consider a relative acceleration with respect to a
reference frame $\G$,  one should prolong the connection $\G$ on
$Q\to\bR$ to a holonomic connection $\xi_\G$ on the jet bundle $J^1Q\to\bR$.
Note that the jet prolongation $J^1\G$ (\ref{gm217'}) of 
$\G$ is not holonomic.
We can construct a desired prolongation by means of a dynamic connection $\g$
on the affine jet bundle $J^1Q\to Q$.

Let us consider the composite bundle $J^1Q\to Q\to\bR$
and connections $\g$ on $J^1Q\to Q$ and $J^1\G$ on
$J^1Q\to\bR$. Then there exists a dynamic connection
$\wt\g$ (\ref{jp10}) on
$J^1Q\to Q$ with the components
\be
\wt \g^i_k=\g^i_k, \qquad \wt \g^i_0=d_t\G^i-\g^i_k\G^k. 
\ee
 Now, let us construct some soldering form and add
it to this connection.  The covariant
derivative of a reference frame $\G$ with respect to the 
dynamic connection $\g$ reads
\beq
\nabla\G= \nabla_\la\G^k dq^\la\ot\dr_k^t: Q\to T^*Q\times V_QJ^1Q, 
\quad \nabla_\la\G^k = \dr_\la\G^k - \g^k_\la\circ\G. \label{jp57}
\eeq
Let us apply the canonical projection $T^*Q\to V^*Q$ and then the imbedding
$\G:V^*Q\to T^*Q$ to (\ref{jp57}).  We obtain the $V_QJ^1Q$-valued 1-form
\be
\si= [-\G^i(\dr_i\G^k - \g^k_i\circ\G)dt +(\dr_i\G^k -
\g^k_i\circ\G)dq^i]\ot\dr_k^t
\ee
on $Q$ whose pull-back onto $J^1Q$ is a desired soldering form. 
The sum $\g_\G=\wt \g +\si$, called the frame connection, reads
\beq
 \g_\G{}^i_0= d_t\G^i - \g^i_k\G^k -\G^k(\dr_k\G^i - \g^i_k\circ\G),\qquad
  \g_\G{}^i_k= \g^i_k +\dr_k\G^i - \g^i_k\circ\G. \label{jp68}
\eeq
This connection yields the holonomic connection
\be
\xi_\G^i= d_t\G^i +(\dr_k\G^i +\g^i_k - \g^i_k\circ\G)(q^k_t-\G^k).
\ee

 Let $\xi$ be a dynamic equation and 
 $\g=\g_\xi$ the connection (\ref{z286}) associated with
$\xi$. Then one can think of the vertical
vector field 
\be
a_\G=\xi-\xi_\G=(\xi^i-\xi_\G^i)\dr_t^i 
\ee
 on the affine jet bundle $J^1Q\to Q$  as being a relative
acceleration with respect to the reference frame $\G$.

For instance, let us consider the reference frame which is geodesic
for the dynamic equation $\xi$, i.e., 
\be
\G\rfloor\nabla \G= (d_t\G^i-\xi^i\circ \G)\dr_i=0, 
\ee
where $\nabla \G$ is the covariant derivative (\ref{jp57}) with respect to
the dynamic connection $\g_\xi$.
It is readily observed that 
integral sections $c$ of a reference frame $\G$ are solutions of a
dynamic equation $\xi$ if and only if $\G$ is the geodesic reference frame
for $\xi$.
Then the relative acceleration of a motion $c$ with respect to the
reference frame $\G$ is $(\xi-\xi_\G)\circ\G=0$.

Let now $\xi$ be an arbitrary dynamic equation, written with respect to
coordinates $(t,q^i)$, proper for the reference frame $\G$, i.e., $\G^i=0$.
The relative acceleration with respect to the frame $\G$ in these
coordinates is
\beq
a^i_\G = \xi^i(t,q^j,q^j_t) -\frac12 q^k_t(\dr_k \xi^i -
\dr_k
\xi^i\mid_{q^j_t=0}). \label{jp64}
\eeq 
Given another bundle coordinates $(t, q'^i)$, 
this dynamic equation reads
\be
&& \xi'^i= \dr_jq'^i\xi^j (t,q^m(t,q'^k),\frac{\dr q^m}{\dr q'^k}(q'^k-\G^k))
+ \\
&& \quad  d_t\G^i +\frac{\dr \G^i}{\dr q'^j}(q'^j_t-\G^j) -
\dr_m q'^i\frac{\dr q^m}{\dr q'^j\dr q'^k}(q'^j_t-\G^j)
(q'^k_t-\G^k), 
\ee
while the relative acceleration (\ref{jp64}) with respect
to the reference frame $\G$ takes the form
\be
a'^i_\G =\dr_jq'^i a^j_\G. 
\ee
Then we can write a dynamic
equation in the form, covariant under coordinate transformations:
\be
\wt D_{\g_\G} q^i_t = d_t q^i_t -\xi^i_\G=a_\G,
\ee
where $\wt D_{\g_\G}$ is the vertical covariant differential (\ref{7.10})
with respect to the frame connection $\g_\G$ (\ref{jp68}) on $J^1Q\to Q$.

In particular, if $\xi$ is a free motion equation which takes the form
(\ref{z280}) with respect to a reference frame $\G$, then
\be
\wt D_{\g_\G} q^i_t=0
\ee 
relative to any coordinates.

\section{Relativistic and non-relativistic dynamic equations}

In physical applications, one usually thinks of non-relativistic mechanics as
being an approximation of small velocities of a relativistic
theory. At the same time, the velocities in mathematical formalism of
non-relativistic mechanics are not bounded. It has long been
recognized that the relation between the mathematical schemes of
relativistic and non-relativistic mechanics is not trivial.  

Let $X$ be a 4-dimensional world manifold of a relativistic theory,
coordinated by
$(x^\la)$. Then the tangent bundle $TX$ of $X$ plays the role of a space of
its 4-velocities.
A relativistic equation of motion is said
to be  a geodesic equation
\be
\dot x^\la\dr_\la\dot x^\m= K_\la^\m(x^\nu,\dot
x^\nu) \dot x^\la
\ee
with respect to a (non-linear) connection $K$ 
on $TX\to X$. 

It is supposed additionally that there is a pseudo-Riemannian metric
$g$ of signature $(+,---)$ in $TX$ such that a geodesic vector field
does not leave the subbundle of
relativistic hyperboloids 
\beq
W_g=\{\dot x^\la\in TX\, \mid \,\,g_{\la\m} \dot x^\la\dot x^\m=1\}
\label{cqg1}
\eeq
in $TX$. It suffices to require that the condition
\beq
(\dr_\la g_{\m\nu}\dot x^\m + 2g_{\m\nu}K^\m_\la)\dot x^\la \dot x^\nu =0.
\label{cqg4}
\eeq
holds for all tangent vectors which belong to $W_g$ (\ref{cqg1}). 
 Obviously, the
Levi--Civita connection 
$\{_\la{}^\m{}_\nu\}$ of the metric $g$ fulfills the
condition (\ref{cqg4}). Any connection
$K$ on $TX\to X$ can be written as
\be
K^\m_\la = \{_\la{}^\m{}_\nu\}\dot x^\nu + \si^\m_\la(x^\la,\dot x^\la),
\ee
where the soldering form $\si=\si^\m_\la dx^\la\ot\dot\dr_\la$ plays the role
of an external force. Then the condition (\ref{cqg4}) takes the form
\beq
g_{\m\nu}\si^\m_\la\dot x^\la \dot x^\nu=0. \label{cqg46}
\eeq

Let now a world manifold $X$ admit a projection $X\to \bR$, where $\bR$ is
a time axis. One can think of the bundle $X\to\bR$ as being a configuration
space of non-relativistic mechanical system. There is the canonical imbedding
(\ref{z260}) of 
$J^1X$ onto the affine subbundle 
\beq
\dot x^0=1, \qquad \dot x^i=x^i_0 \label{cqg7}
\eeq
of the tangent
bundle $TX$.
Then one can think of (\ref{cqg7}) as the 4-velocities of a
non-relativistic system. The relation (\ref{cqg7}) differs from the
familiar relation  between 4- and 3-velocities of a relativistic
system. In particular, the
temporal component
$\dot x^0$ of 4-velocities of a non-relativistic system equals 1 (relative to
the universal unit system). It follows that the 4-velocities of
relativistic and non-relativistic systems occupy different subbundles of the
tangent bundle
$TX$. Moreover, Theorem \ref{jp50} shows
that both relativistic and non-relativistic equations of
motion can be seen as the geodesic equations on the same tangent bundle
$TX$, but their solutions
live in the different subbundles (\ref{cqg1}) and (\ref{cqg7}) of $TX$.  
At the same time, relativistic equations, expressed into the  3-velocities
$\dot x^i/\dot x^0$ of a relativistic system, tend exactly to the
non-relativistic  equations on the subbundle (\ref{cqg7}) when $\dot x^0\to
1$, $g_{00}\to 1$, i.e., where non-relativistic mechanics and the
non-relativistic approximation of a relativistic theory coincide only.

Given a coordinate systems $(x^0,x^i)$, compatible with the fibration $X\to
\bR$, let us consider a non-degenerate quadratic Lagrangian 
\beq
L=\frac12m_{ij}(x^\m) x^i_0 x^j_0 + k_i(x^\m) x^i_0  +
f(x^\m), \label{cqg20}
\eeq
where $m_{ij}$ is a Riemannian mass tensor. Similarly to
Proposition
\ref{aff}, one can show that any quadratic polynomial in $J^1X\subset TX$ is
extended to a bilinear form in $TX$. Then the Lagrangian $L$
(\ref{cqg20}) can be written as 
\beq
L=-\frac12g_{\al\m}x^\al_0 x^\m_0, \qquad x^0_0=1, \label{cqg40}
\eeq
where $g$ is the metric
\beq
g_{00}=-2f, \qquad g_{0i}=-k_i, \qquad g_{ij}=-m_{ij}. \label{cqg21}
\eeq
The corresponding Lagrange equation takes the form
\beq
x^i_{00}=-(m^{-1})^{ik}\{_{\la k\nu}\}x^\la_0x^\nu_0, \qquad x^0_0=1,
\label{cqg35}
\eeq
where 
\be
\{_{\la\m\nu}\} =-\frac12(\dr_\la g_{\m\nu} +\dr_\nu
g_{\m\la} - \dr_\m g_{\la\nu})
\ee
 are the Christoffel symbols of the metric (\ref{cqg21}). Let us assume that
this metric is non-degenerate. By virtue of
Corollary \ref{c2}, the dynamic equation (\ref{cqg35}) gives rise to
the geodesic equation on $TX$ 
\be
&& \dot x^\la\dr_\la \dot x^0 = 0, \qquad \dot x^0=1,  \\
&&\dot x^\la\dr_\la \dot x^i = \{_\la{}^i{}_\nu\}\dot x^\la\dot x^\nu -
g^{i0}\{_{\la 0\nu}\}\dot x^\la\dot x^\nu. 
\ee

Let us now bring the Lagrangian (\ref{cqg20}) into the form
\beq
L=\frac12m_{ij}(x^\m)( x^i_0-\G^i)(x^j_0-\G^j) +
f'(x^\m), \label{cqg48}
\eeq
where $\G$ is a Lagrangian connection on $X\to\bR$. This connection $\G$
defines an atlas of local constant trivializations of the
bundle $X\to\bR$ and the corresponding coordinates $(x^0,\ol x^i)$ on $X$. In
this coordinates, the Lagrangian $L$ (\ref{cqg48}) reads
\beq
L=\frac12\ol m_{ij}\ol x^i_0 \ol x^j_0 +
f'(x^\m). \label{jp73}
\eeq
One can think of its first term as the kinetic energy of
a non-relativistic system with the mass tensor $\ol m_{ij}$ relative to the
reference frame
$\G$, while $(-f')$ is a potential.  Let us assume that
$f'$ is a nowhere vanishing function on
$X$. Then the Lagrange equation (\ref{cqg35}) takes the form
\beq
\ol x^i_{00}=\{_\la{}^i{}_\nu\}\ol x^\la_0\ol x^\nu_0, \qquad \ol x^0_0=1,
\label{jp71}
\eeq
where $\{_\la{}^i{}_\nu\}$ are the Christoffel symbols of the metric
\beq
g_{ij}= -\ol m_{ij}, \qquad g_{0i}=0, \qquad g_{00} =-2f'. \label{cqg51}
\eeq
This metric is Riemannian if $f'>0$ and pseudo-Riemannian if $f'<0$.
 Then the spatial part of the corresponding geodesic
equation
\ben
&& \dot {\ol x}^\la\dr_\la \dot{\ol x}^0 = 0, \qquad \dot{\ol x}^0=1,
\nonumber \\ 
&&\dot {\ol x}^\la\dr_\la \dot {\ol x}^i =
\{_\la{}^i{}_\nu\}\dot{\ol x}^\la\dot{\ol x}^\nu \label{jp70}
\een
is exactly the spatial part of the geodesic equation with
respect to the Levi--Civita connection of the metric (\ref{cqg51}) on $TX$.
It follows that  the
non-relativistic dynamic equation (\ref{jp71}) describes the
non-relativistic approximation of the geodesic motion in a curved space with
the metric (\ref{cqg51}).

Conversely, let us consider a geodesic motion
\beq
\dot x^\la\dr_\la \dot x^\m=\{_\la{}^\m{}_\nu\}\dot x^\la\dot x^\nu
\label{cqg70}
\eeq
in the presence of a
pseudo-Riemannian metric $g$ on a world manifold $X$. Let $(x^0,\ol x^i)$ be
local hyperbolic coordinates such that $g_{00}=1$, $g_{0i}=0$. These
coordinates set a non-relativistic reference frame for a local fibration
$X\to\bR$. Then the equation (\ref{cqg70}) has the non-relativistic limit
(\ref{jp70}) which is the Lagrange equation for the Lagrangian (\ref{jp73})
where $f'=0$. This Lagrangian 
describes a free 
non-relativistic mechanical system with the mass tensor $\ol m_{ij}=-g_{ij}$.

In view of Proposition \ref{jp40}, the "relativization"
(\ref{cqg40}) of an arbitrary non-relativistic quadratic Lagrangian
(\ref{cqg20}) may lead to a confusion. In particular, it can be applied to a
gravitational Lagrangian (\ref{cqg48}) where $f'$ is a gravitational
potential. An arbitrary quadratic dynamic equation can be written in the form
\be
x^i_{00}=-(m^{-1})^{ik}\{_{\la k\m}\} x^\la_0 x^\m_0 + b^i_\m(x^\nu)x^\m_0,
\qquad x^0_0=1,
\ee
where $\{_{\la k\m}\}$ are the Christoffel symbols of some
 pseudo-Riemannian metric $g$, whose spatial part is the mass tensor
$(-m_{ik})$, while
\beq
b^i_k(x^\m) x^k_0 + b^i_0(x^\m) \label{cqg61}
\eeq
is an external force. With respect to the coordinates where  $g_{0i}=0$, one
may construct the relativistic equation
\beq
\dot x^\la\dr_\la\dot x^\m= \{_\la{}^\m{}_\nu\}\dot x^\la\dot x^\nu +
\si^\m_\la \dot x^\la, \label{cqg73}
\eeq
where the soldering form $\si$ must fulfill the condition (\ref{cqg46}). It
takes place only if 
\be
g_{ik}b^i_j + g_{ij}b^i_k=0,
\ee
i.e., the external force (\ref{cqg61}) is the Lorentz-type force plus 
some potential one. Then, we have
\be
\si^0_0= 0, \qquad \si^0_k = -g^{00}g_{kj}b^j_0, \qquad \si^j_k=b^j_k.
\ee

The "relativization" (\ref{cqg73}) exhausts almost all familiar examples.
It means that a wide class of mechanical system can be represented as a
geodesic motion with respect to some affine connection in the spirit of
Cartan's idea.

\end{document}